\documentclass[letterpaper, 10 pt, conference]{ieeeconf}
\usepackage{amsmath,amsfonts,amssymb,mathrsfs}
\usepackage{graphicx}
\usepackage{BernsteinStyle}
\usepackage{mathtools}
\usepackage{booktabs}
\usepackage{hyperref}
\usepackage{cite}
\usepackage{placeins}

\newcommand{\papergeometryscale}{0.92}
\AtBeginDocument{%
  \setlength{\floatsep}{\papergeometryscale\floatsep}%
  \setlength{\textfloatsep}{\papergeometryscale\textfloatsep}%
  \setlength{\dblfloatsep}{\papergeometryscale\dblfloatsep}%
  \setlength{\dbltextfloatsep}{\papergeometryscale\dbltextfloatsep}%
  \setlength{\intextsep}{\papergeometryscale\intextsep}%
  \setlength{\abovedisplayskip}{\papergeometryscale\abovedisplayskip}%
  \setlength{\belowdisplayskip}{\papergeometryscale\belowdisplayskip}%
  \setlength{\abovedisplayshortskip}{\papergeometryscale\abovedisplayshortskip}%
  \setlength{\belowdisplayshortskip}{\papergeometryscale\belowdisplayshortskip}%
  \setlength{\abovecaptionskip}{\papergeometryscale\abovecaptionskip}%
  \setlength{\belowcaptionskip}{\papergeometryscale\belowcaptionskip}%
  \setlength{\jot}{\papergeometryscale\jot}%
}

\makeatletter
\renewcommand\section{\@startsection{section}{1}{\z@}{1.5ex}{0.7ex}%
  {\normalfont\normalsize\centering\scshape}}
\renewcommand\subsection{\@startsection{subsection}{2}{\z@}{1.5ex}{0.7ex}%
  {\normalfont\normalsize\itshape}}
\makeatother

\IEEEoverridecommandlockouts

\title{\LARGE \bf
Taylor-Informed Predictive Cost Adaptive Control \\ for Quadrotors with Online Gravity-Trim Adaptation
}

\author{Tam W. Nguyen$^{1}$
\thanks{$^{1}$Tam W. Nguyen is with the Department of Electrical Engineering, Kyoto University, Kyoto 615-8510, Japan {\tt\small nguyen.tamwilly.3e@kyoto-u.ac.jp}}%
}

\begin{document}

\maketitle
\thispagestyle{empty}
\pagestyle{empty}

\begin{abstract}
This paper develops Taylor-informed predictive cost adaptive control
(PCAC) for quadrotors with online gravity-trim adaptation. First-, second-,
and third-order expansions of the nonlinear dynamics about nominal hover
define sparse sampled-data dictionaries for row-wise recursive
least-squares identification with variable-rate forgetting. At each step,
the identified predictor is linearized at the current state, and its
Jacobian is fixed over the prediction horizon. The identified vertical
dynamics also estimate the vehicle mass and gravity-trim input, eliminating
fixed nominal gravity compensation. Simulations with an abrupt payload
change and aggressive helix tracking show that the higher-order predictors
improve prediction and tracking while preserving the standard PCAC
formulation.
\end{abstract}

\section{Introduction}

Quadrotors are a representative nonlinear control problem because they
combine underactuation, strong coupling, and real-time constraints
\cite{lee2010geometric,rastgoftar2021safe,hermand2018constrained}.
Model uncertainty, payload changes, and disturbances motivate adaptive
flight-control methods that maintain performance without repeated offline
identification
\cite{dydek2013adaptive,richards2025experimental,he2026projection,
vanderschaaf2026active}.

Model predictive control (MPC) provides a systematic treatment of tracking
objectives and constraints \cite{eren2017model}. Adaptive MPC combines
predictive optimization with online identification
\cite{clarke1987generalized,clarke1987generalized2,tao2026robust}.
Predictive cost adaptive control (PCAC), in particular, identifies a
sampled-data predictor and constructs the local prediction model directly
from the identified dynamics
\cite{nguyen2021predictive,richards2025benchmark,farahmandi2024missile}.
Related work has extended adaptive prediction to nonlinear and behavioral
models
\cite{nguyen2026adaptivebehavioralpredictivecontrol,
alhazmi2026nonlinearpredictivecostadaptive}.

Payload changes alter the vehicle mass and corresponding hover thrust. A
fixed nominal gravity-compensation term therefore produces steady-state
error when the true mass differs from its nominal value. Although adaptive
flight control can estimate changing parameters
\cite{lai2024adaptive,burtsev2026adaptive,huang2026fls}, existing quadrotor
PCAC relies on a sparse hover-linear predictor \cite{nguyen2026fast}, whose
prediction accuracy degrades during larger excursions from hover.

This paper develops a nonlinear formulation based on local Taylor structure.
First-, second-, and third-order expansions about nominal hover
generate sparse sampled-data dictionaries for row-wise recursive
least-squares identification. The identified vertical dynamics provide
online mass and gravity-trim estimates. At each control step, the nonlinear
predictor is linearized at the current state and held fixed over the horizon,
yielding an affine model for the standard PCAC optimization.

Simulations evaluate an abrupt payload change and aggressive helix
tracking. Higher-order predictors improve tracking and one-step prediction,
and the trim estimate removes the need for fixed nominal compensation. The
complete MATLAB implementation is available at
\url{https://github.com/tamwng/taylor-informed-pcac-quadrotor}.

\section{Problem Setup}\label{sec:2}

Let $\Sigma_\rmi$ and $\Sigma_\rmb$ denote the inertial and body-fixed
frames, respectively, and let $e_j\in\mathbb R^3$, $j\in\{1,2,3\}$,
be the Euclidean basis vectors. The inertial $e_3$-axis points upward.
The position and velocity in $\Sigma_\rmi$ are $p\in\mathbb R^3$ and
$v\coloneqq\dot p$. The attitude
$R\coloneqq{}^\rmi R_\rmb\in SO(3)$ maps body-frame components to
inertial coordinates, and $\omega\in\mathbb R^3$ is the body-frame
angular velocity. Let $f\geq0$ be the total thrust along the body
$e_3$-axis and $\tau\in\mathbb R^3$ the body torque. The dynamics are
\begin{align}
        \dot p &= v, \label{eq:p_dot} \\
        m \dot v &= - m g e_3 + f R e_3, \label{eq:v_dot} \\
        \dot R &= R \omega^\times, \label{eq:R_dot} \\
        J \dot \omega &= - \omega^\times J \omega + \tau,
        \label{eq:omega_dot}
\end{align}
where $m>0$, $J=J^\top\succ0$, and $g>0$ denote the mass, inertia,
and gravitational acceleration. For a body-frame vector $a_\rmb$,
$a_\rmi=Ra_\rmb$.

For $\omega = [\omega_1 \ \omega_2 \ \omega_3]^\top$, the
skew-symmetric map $(\cdot)^\times : \mathbb R^3 \to
\mathfrak{so}(3)$ is defined by
\begin{align}
        \omega^\times \coloneqq
        \left[\begin{smallmatrix}
        0 & -\omega_3 & \omega_2 \\
        \omega_3 & 0 & -\omega_1 \\
        -\omega_2 & \omega_1 & 0
        \end{smallmatrix}\right].
\end{align}

Parameterize the attitude by the intrinsic $Z$-$X$-$Y$ angles
$\xi\coloneqq
[\begin{smallmatrix}\psi&\varphi&\theta\end{smallmatrix}]^\top$,
where $\psi$, $\varphi$, and $\theta$ are yaw, roll, and pitch. Then
$R(\xi)=R_z(\psi)R_x(\varphi)R_y(\theta)$.
Using $c_\alpha \coloneqq \cos \alpha$ and
$s_\alpha \coloneqq \sin \alpha$ yields
\begin{align}
R(\xi) \hspace{-0.2em}=\hspace{-0.2em}
\left[\begin{smallmatrix}
c_\psi c_\theta - s_\varphi s_\psi s_\theta
&
- s_\psi c_\varphi
&
c_\psi s_\theta + s_\varphi s_\psi c_\theta
\\
s_\psi c_\theta + s_\varphi c_\psi s_\theta
&
c_\psi c_\varphi
&
s_\psi s_\theta - s_\varphi c_\psi c_\theta
\\
- c_\varphi s_\theta
&
s_\varphi
&
c_\varphi c_\theta
\end{smallmatrix}\right].
\end{align}

The angular velocity satisfies $\omega=E(\xi)\dot\xi$, where
\begin{align}
E(\xi) &=
\left[\begin{smallmatrix}
- s_\theta c_\varphi & c_\theta & 0 \\
s_\varphi & 0 & 1 \\
c_\theta c_\varphi & s_\theta & 0
\end{smallmatrix}\right].
\end{align}
Hence, for $c_\varphi\neq0$, $\dot\xi=E^{-1}(\xi)\omega$, where
\begin{align}
E^{-1}(\xi) &=
\left[\begin{smallmatrix}
-\frac{s_\theta}{c_\varphi} & 0 & \frac{c_\theta}{c_\varphi}
\\[2mm]
c_\theta & 0 & s_\theta
\\[1mm]
s_\theta \tan \varphi & 1 & -c_\theta \tan \varphi
\end{smallmatrix}\right].
\end{align}
Define the state and input as
\begin{align}
        x &\coloneqq
        \begin{bmatrix}
        p^\top & \xi^\top & v^\top & \omega^\top
        \end{bmatrix}^\top
        \in \mathbb R^{12}, \\
        u &\coloneqq
        \begin{bmatrix}
        f & \tau^\top
        \end{bmatrix}^\top
        \in \mathbb R^4 .
\end{align}
Equations \eqref{eq:p_dot}--\eqref{eq:omega_dot} become
\begin{align}\label{eq:sys_exact1}
        \dot x = \mathcal F(x,u;m,J),
\end{align}
where 
\begin{align}
\mathcal F(x,u;m,J)
&=
\begin{bmatrix}
v
\\[1mm]
E^{-1}(\xi)\omega
\\[1mm]
- g e_3 + \dfrac{f}{m} R(\xi)e_3
\\[2mm]
J^{-1}\left(-\omega^\times J\omega+\tau\right)
\end{bmatrix}. \label{eq:sys_exact2}
\end{align}

\section{Nominal Hover Coordinates and Sampled Dynamics}\label{sec:3}

For any $p_\rme\in\mathbb R^3$, the hover equilibrium for mass $m$ is
\begin{align}
        x_\rme &=
        \begin{bmatrix}
        p_\rme^\top & 0_{1\times 3} & 0_{1\times 3} & 0_{1\times 3}
        \end{bmatrix}^\top, \\
        u_\rme(m) &=
        \begin{bmatrix}
        mg & 0 & 0 & 0
        \end{bmatrix}^\top .
\end{align}
For nominal mass $m_0$, set
$u_{\rme,0} \coloneqq
        [\begin{smallmatrix}m_0 g & 0 & 0 & 0\end{smallmatrix}]^\top$.
The pair $(x_\rme,u_{\rme,0})$ is not an equilibrium when $m\neq m_0$.

Define the shifted variables
$\tilde x \coloneqq x - x_\rme$, \qquad
$\tilde u \coloneqq u - u_{\rme,0}$.
The shifted dynamics are
\begin{align}
        \dot{\tilde x}
        &=
        \bar{\mathcal F}(\tilde x,\tilde u;m,m_0,J) \notag \\
        &\coloneqq
        \mathcal F(x_\rme+\tilde x,u_{\rme,0}+\tilde u;m,J).
\end{align}
At the nominal hover point,
\begin{align}
        \bar{\mathcal F}(0,0;m,m_0,J)
        =
        \left[\begin{smallmatrix}
        0_{3\times 1}
        \\
        0_{3\times 1}
        \\
        g\left(\frac{m_0}{m}-1\right)e_3
        \\
        0_{3\times 1}
        \end{smallmatrix}\right].
\end{align}
The mass mismatch therefore appears as a constant vertical acceleration.

Under zero-order hold with sampling period $T_s>0$,
$u(t) = u_k$, $t \in [kT_s,(k+1)T_s)$,
denote the flow map of $\mathcal F(\cdot,\cdot;m,J)$ by $F_{m,J}$:
\begin{align}
        x_{k+1} = F_{m,J}(x_k,u_k).\label{eq:DT1}
\end{align}
With full-state feedback, the shifted sampled-data dynamics are
\begin{align}
        \tilde x_{k+1}
        =
        \bar F_{m,m_0,J}(\tilde x_k,\tilde u_k),\label{eq:DT2}
\end{align}
where
\begin{align}
        \bar F_{m,m_0,J}(\tilde x,\tilde u)
        \coloneqq
        F_{m,J}(x_\rme+\tilde x,u_{\rme,0}+\tilde u)-x_\rme .\label{eq:DT3}
\end{align}
Define $\beta_g(m,m_0)
        \coloneqq
        g\left(\frac{m_0}{m}-1\right).$
At $(\tilde x,\tilde u)=(0,0)$, the sampled-data residual is
\begin{align}
        \bar F_{m,m_0,J}(0,0)
        =
        \left[\begin{smallmatrix}
        \frac{T_s^2}{2}\beta_g(m,m_0)e_3
        \\
        0_{3\times 1}
        \\
        T_s\beta_g(m,m_0)e_3
        \\
        0_{3\times 1}
        \end{smallmatrix}\right].
\end{align}
Consequently,
\begin{align}
        \bar F_{m,m_0,J}(0,0)=0
        \quad \Longleftrightarrow \quad
        m=m_0 .
\end{align}

\section{Taylor Expansion About the Nominal Hover Point}\label{sec:4}

Partition the shifted input as
\begin{align}
        \tilde u
        =
        \begin{bmatrix}
        \tilde f & \tilde \tau^\top
        \end{bmatrix}^\top ,
        \qquad
        \tilde \tau =
        \begin{bmatrix}
        \tilde \tau_1 & \tilde \tau_2 & \tilde \tau_3
        \end{bmatrix}^\top .
        \label{eq:shifted_input_taylor}
\end{align}
Because the nominal torque is zero, $\tilde\tau=\tau$, while
$f=m_0g+\tilde f$.

Define
$\alpha_0(m,m_0)
        \coloneqq
        \frac{m_0 g}{m},
        \alpha_f(m)
        \coloneqq
        \frac{1}{m}$,
and let
$z \coloneqq
        [\begin{smallmatrix}\tilde x^\top & \tilde u^\top\end{smallmatrix}]^\top$.
Expand about $z=0$, corresponding to $(x_\rme,u_{\rme,0})$. The Euler-angle
representation is local to a neighborhood with $c_\varphi\neq0$.

For $D\in\{1,2,3\}$, let $\bar{\mathcal F}^{[D]}$ be the degree-$D$
Taylor polynomial of $\bar{\mathcal F}$ at $z=0$:
\begin{align}
        \bar{\mathcal F}(\tilde x,\tilde u;m,m_0,J)
        =
        \bar{\mathcal F}^{[D]}(\tilde x,\tilde u;m,m_0,J)
        +
        r_D(z),
        \label{eq:ct_taylor_remainder}
\end{align}
where $r_D(z)$ contains terms of total degree greater than $D$. Write the
retained field as
\begin{align}
        \bar{\mathcal F}^{[D]}(\tilde x,\tilde u;m,m_0,J)
        =
        \begin{bmatrix}
        v
        \\
        \eta_D(\xi,\omega)
        \\
        a_D(\xi,\tilde f;m,m_0)
        \\
        \Omega_D(\omega,\tilde \tau;J)
        \end{bmatrix}.
        \label{eq:hover_taylor_vector_field}
\end{align}
The position row is exact; the other rows depend on $D$.

The Euler-kinematic terms are
\begin{align}
        \eta_1(\xi,\omega)
        &=
        \left[\begin{smallmatrix}
        \omega_3 \\
        \omega_1 \\
        \omega_2
        \end{smallmatrix}\right],
        \label{eq:eta_1}
        \\
        \eta_2(\xi,\omega)
        &=
        \eta_1(\xi,\omega)
        +
        \left[\begin{smallmatrix}
        -\theta \omega_1 \\
        \theta \omega_3 \\
        -\varphi \omega_3
        \end{smallmatrix}\right],
        \label{eq:eta_2}
        \\
        \eta_3(\xi,\omega)
        &=
        \eta_2(\xi,\omega)
        +
        \left[\begin{smallmatrix}
        \frac{1}{2}(\varphi^2-\theta^2)\omega_3 \\
        -\frac{1}{2}\theta^2\omega_1 \\
        \theta\varphi\omega_1
        \end{smallmatrix}\right].
        \label{eq:eta_3}
\end{align}

The translational acceleration terms are
\begin{align}
        a_1(\xi,\tilde f;m,m_0)
        &=
        \left[\begin{smallmatrix}
        \alpha_0 \theta
        \\
        -\alpha_0 \varphi
        \\
        \beta_g+\alpha_f\tilde f
        \end{smallmatrix}\right],
        \label{eq:a_1}
        \\
        a_2(\xi,\tilde f;m,m_0)
        &=
        a_1(\xi,\tilde f;m,m_0)
        +
        \hspace{-0.2em}\left[\begin{smallmatrix}
        \alpha_0\varphi\psi+\alpha_f\tilde f\theta
        \\
        \alpha_0\psi\theta-\alpha_f\tilde f\varphi
        \\
        -\frac{\alpha_0}{2}(\varphi^2+\theta^2)
        \end{smallmatrix}\right],
        \label{eq:a_2}
        \\
        a_3(\xi,\tilde f;m,m_0)
        &=
        a_2(\xi,\tilde f;m,m_0)
        \nonumber
        \\
        &\quad+
        \left[\begin{smallmatrix}
        -\frac{\alpha_0}{6}\theta^3
        -\frac{\alpha_0}{2}\psi^2\theta
        +\alpha_f\tilde f\varphi\psi
        \\
        \frac{\alpha_0}{6}\varphi^3
        +\frac{\alpha_0}{2}\varphi\psi^2
        +\frac{\alpha_0}{2}\varphi\theta^2
        +\alpha_f\tilde f\psi\theta
        \\
        -\frac{\alpha_f}{2}\tilde f(\varphi^2+\theta^2)
        \end{smallmatrix}\right].
        \label{eq:a_3}
\end{align}
Arguments of $\alpha_0$, $\alpha_f$, and $\beta_g$ are suppressed in
\eqref{eq:a_1}--\eqref{eq:a_3}.

The angular-acceleration terms are
\begin{align}
        \Omega_1(\omega,\tilde \tau;J)
        &=
        J^{-1}\tilde \tau,
        \label{eq:Omega_1}
        \\
        \Omega_2(\omega,\tilde \tau;J)
        &=
        \Omega_3(\omega,\tilde \tau;J)
        =
        J^{-1}\left(-\omega^\times J\omega+\tilde \tau\right).
        \label{eq:Omega_23}
\end{align}
For principal-axis body coordinates,
$J=\operatorname{diag}(J_1,J_2,J_3)$ and
\begin{align}
        \Omega_2(\omega,\tilde \tau;J)
        =
        \Omega_3(\omega,\tilde \tau;J)
        =
        \left[\begin{smallmatrix}
        \frac{J_2-J_3}{J_1}\omega_2\omega_3
        +\frac{1}{J_1}\tilde \tau_1
        \\[2mm]
        \frac{J_3-J_1}{J_2}\omega_3\omega_1
        +\frac{1}{J_2}\tilde \tau_2
        \\[2mm]
        \frac{J_1-J_2}{J_3}\omega_1\omega_2
        +\frac{1}{J_3}\tilde \tau_3
        \end{smallmatrix}\right].
        \label{eq:Omega_diag}
\end{align}

\section{Sparse Taylor-Informed Sampled-Data Dictionary}\label{sec:dictionary}

For $J=\operatorname{diag}(J_1,J_2,J_3)$, the Taylor terms in
\eqref{eq:eta_1}--\eqref{eq:Omega_23} define a row-wise sampled-data
dictionary. Table~\ref{tab:regressor_D1} lists the complete degree-one
regressors; Table~\ref{tab:regressor_D23} lists the additional degree-two
and degree-three monomials.

\begin{table}[t]
\centering
\caption{Complete degree-one row regressors.}
\scriptsize
\begin{tabular}{cl}
\toprule
State row $q$ & $\phi_{q,1}(z)$\\
\midrule
$\tilde p_1$
&
$[\tilde p_1\;\;v_1\;\;\theta\;\;\omega_2\;\;\tilde\tau_2]^\top$
\\
$\tilde p_2$
&
$[\tilde p_2\;\;v_2\;\;\varphi\;\;\omega_1\;\;\tilde\tau_1]^\top$
\\
$\tilde p_3$
&
$[\tilde p_3\;\;v_3\;\;1\;\;\tilde f]^\top$
\\
$\psi$
&
$[\psi\;\;\omega_3\;\;\tilde\tau_3]^\top$
\\
$\varphi$
&
$[\varphi\;\;\omega_1\;\;\tilde\tau_1]^\top$
\\
$\theta$
&
$[\theta\;\;\omega_2\;\;\tilde\tau_2]^\top$
\\
$v_1$
&
$[v_1\;\;\theta\;\;\omega_2\;\;\tilde\tau_2]^\top$
\\
$v_2$
&
$[v_2\;\;\varphi\;\;\omega_1\;\;\tilde\tau_1]^\top$
\\
$v_3$
&
$[v_3\;\;1\;\;\tilde f]^\top$
\\
$\omega_1$
&
$[\omega_1\;\;\tilde\tau_1]^\top$
\\
$\omega_2$
&
$[\omega_2\;\;\tilde\tau_2]^\top$
\\
$\omega_3$
&
$[\omega_3\;\;\tilde\tau_3]^\top$
\\
\bottomrule
\end{tabular}
\label{tab:regressor_D1}
\end{table}

\begin{table}[t]
\centering
\caption{Additional Taylor monomials retained at degrees two and three.}
\label{tab:regressor_D23}
\scriptsize
\setlength{\tabcolsep}{4pt}
\renewcommand{\arraystretch}{1.05}
\resizebox{\columnwidth}{!}{%
\begin{tabular}{ccc}
\toprule
State row $q$
&
Degree-two addition $\delta\phi_{q,2}(z)$
&
Degree-three addition $\delta\phi_{q,3}(z)$
\\
\midrule
$\tilde p_1,\;v_1$
&
$[\varphi\psi,\;\tilde f\theta]^\top$
&
$[\theta^3,\;\psi^2\theta,\;\tilde f\varphi\psi]^\top$
\\
$\tilde p_2,\;v_2$
&
$[\psi\theta,\;\tilde f\varphi]^\top$
&
$[\varphi^3,\;\varphi\psi^2,\;\varphi\theta^2,\;\tilde f\psi\theta]^\top$
\\
$\tilde p_3,\;v_3$
&
$[\varphi^2,\;\theta^2]^\top$
&
$[\tilde f\varphi^2,\;\tilde f\theta^2]^\top$
\\
$\psi$
&
$[\theta\omega_1]^\top$
&
$[\varphi^2\omega_3,\;\theta^2\omega_3]^\top$
\\
$\varphi$
&
$[\theta\omega_3]^\top$
&
$[\theta^2\omega_1]^\top$
\\
$\theta$
&
$[\varphi\omega_3]^\top$
&
$[\varphi\theta\omega_1]^\top$
\\
$\omega_1$
&
$[\omega_2\omega_3]^\top$
&
$\varnothing$
\\
$\omega_2$
&
$[\omega_3\omega_1]^\top$
&
$\varnothing$
\\
$\omega_3$
&
$[\omega_1\omega_2]^\top$
&
$\varnothing$
\\
\bottomrule
\end{tabular}
}
\end{table}

For each row, $\phi_{q,D}$ concatenates the table entries through degree
$D$, removes duplicates while preserving their first occurrence, and omits
empty entries. Translational-acceleration monomials appear in both the
position and velocity rows.

For example, the vertical-velocity row uses
\begin{align}
        \phi_{v_3,1}(z)
        &=
        \begin{bmatrix}
        v_3 & 1 & \tilde f
        \end{bmatrix}^\top ,
        \\
        \phi_{v_3,2}(z)
        &=
        \begin{bmatrix}
        v_3 & 1 & \tilde f &
        \varphi^2 & \theta^2
        \end{bmatrix}^\top ,
        \\
        \phi_{v_3,3}(z)
        &=
        \begin{bmatrix}
        v_3 & 1 & \tilde f &
        \varphi^2 & \theta^2 &
        \tilde f\varphi^2 & \tilde f\theta^2
        \end{bmatrix}^\top .
        \label{eq:vertical_velocity_regressors}
\end{align}

Order the state rows as
\begin{align}
        \mathcal Q
        \coloneqq
        (q_1,\ldots,q_{12})
        =
        (&\tilde p_1,\tilde p_2,\tilde p_3,
        \psi,\varphi,\theta,
        \nonumber\\
        &v_1,v_2,v_3,
        \omega_1,\omega_2,\omega_3).
        \label{eq:ordered_state_rows}
\end{align}
For each $q\in\mathcal Q$ and $D\in\{1,2,3\}$, define $n_{q,D}
        \coloneqq
        \dim\phi_{q,D}$.
The identified row model is
\begin{align}
        \hat{\bar F}_{q,D,k}(z)
        =
        \hat\Theta_{q,D,k}^\top\phi_{q,D}(z),
        \qquad
        \hat\Theta_{q,D,k}\in\mathbb R^{n_{q,D}}.
        \label{eq:row_sampled_predictor}
\end{align}
Stacking the row models gives
\begin{align}
        \hat{\bar F}_{D,k}(z)
        \coloneqq
        \operatorname{col}_{i=1}^{12}
        \left(
        \hat{\bar F}_{q_i,D,k}(z)
        \right).
        \label{eq:sampled_predictor_collected}
\end{align}

The three dictionaries contain
\begin{align}
        \sum_{q\in\mathcal Q}n_{q,1}
        &=40, \\
        \sum_{q\in\mathcal Q}n_{q,2}
        &=58, \\
        \sum_{q\in\mathcal Q}n_{q,3}
        &=80.
        \label{eq:dictionary_parameter_counts}
\end{align}
Degree one is the sparse hover-linear model with a vertical trim intercept.
Degree two adds thrust--attitude, Euler-kinematic, and gyroscopic couplings;
degree three adds cubic thrust-direction and Euler-kinematic terms.

\section{Taylor-Informed Recursive Identification}\label{sec:6}

Apply RLS independently to each state row. For fixed
$D\in\{1,2,3\}$ and $q\in\mathcal Q$, define
\begin{align}
        \varphi_{q,D,k}
        \coloneqq
        \phi_{q,D}(z_k)
        \in
        \mathbb R^{n_{q,D}}.
        \label{eq:rls_regressor}
\end{align}
The one-step prediction
\begin{align}
        \hat{\tilde x}_{q,k+1|k}
        =
        \hat\Theta_{q,D,k}^{\top}
        \varphi_{q,D,k},
        \label{eq:rls_prediction}
\end{align}
has error
\begin{align}
        e_{q,D,k+1}
        \coloneqq
        \tilde x_{q,k+1}
        -
        \hat{\tilde x}_{q,k+1|k}.
        \label{eq:rls_error}
\end{align}

Initialize
\begin{align}
        \hat\Theta_{q,D,0}
        &\in
        \mathbb R^{n_{q,D}},
        \\
        P_{q,D,0}
        &=
        p_{q,0}I_{n_{q,D}},
        \qquad
        p_{q,0}>0.
        \label{eq:rls_initialization}
\end{align}
For each transition $(z_k,\tilde x_{k+1})$, update
\begin{align}
        L_{q,D,k}
        &=
        \lambda_k^{-1}P_{q,D,k},
        \\
        P_{q,D,k+1}
        &=
        L_{q,D,k}
        -
        L_{q,D,k}\varphi_{q,D,k}
        \nonumber\\
        &\quad\times
        \left(
        1+
        \varphi_{q,D,k}^{\top}
        L_{q,D,k}
        \varphi_{q,D,k}
        \right)^{-1}
        \varphi_{q,D,k}^{\top}
        L_{q,D,k},
        \label{eq:rls_covariance}
        \\
        \hat\Theta_{q,D,k+1}
        &=
        \hat\Theta_{q,D,k}
        +
        P_{q,D,k+1}
        \varphi_{q,D,k}
        e_{q,D,k+1},
        \label{eq:rls_parameter}
\end{align}
The forgetting factor $\lambda_k\in(0,1]$ is computed from the stacked error
\begin{align}
        e_{D,k}
        \coloneqq
        \operatorname{col}_{q\in\mathcal Q}
        \left(
        e_{q,D,k}
        \right)
\end{align}
using the variable-rate forgetting (VRF) rule in
\cite{bruce2020convergence}; $\lambda_k=1$ disables forgetting.

Let $\hat c_{D,k}$ and $\hat b_{D,k}$ be the coefficients of the constant
and $\tilde f$ in the vertical-velocity row. They yield
\begin{align}
        \hat\beta_{g,k}
        &=
        \frac{\hat c_{D,k}}{T_s},
        \\
        \hat\alpha_{f,k}
        &=
        \frac{\hat b_{D,k}}{T_s},
        \\
        \hat m_k
        &=
        \frac{1}{\hat\alpha_{f,k}},
        \\
        \hat{\tilde f}_{\mathrm{trim},k}
        &=
        -
        \frac{\hat\beta_{g,k}}
             {\hat\alpha_{f,k}},
        \\
        \hat f_{\mathrm{trim},k}
        &=
        m_0g+\hat{\tilde f}_{\mathrm{trim},k}.
        \label{eq:trim_estimation}
\end{align}
Project $\hat\alpha_{f,k}$ onto a positive admissible interval before
computing $\hat m_k$ and $\hat f_{\mathrm{trim},k}$.

\section{Predictive Cost Adaptive Control}\label{sec:7}

For fixed $D\in\{1,2,3\}$, linearize the identified map at the current
operating point:
\begin{align}
        z_k
        &\coloneqq
        \begin{bmatrix}
        \tilde x_k^\top & \tilde u_k^\top
        \end{bmatrix}^\top ,
        \\
        \left.
        \frac{\partial\hat{\bar F}_{D,k}}{\partial z}
        \right|_{z=z_k}
        &=
        \begin{bmatrix}
        A_{D,k} & B_{D,k}
        \end{bmatrix},
        \\
        c_{D,k}
        &\coloneqq
        \hat{\bar F}_{D,k}(z_k)
        -
        A_{D,k}\tilde x_k
        -
        B_{D,k}\tilde u_k .
        \label{eq:frozen_predictor}
\end{align}
For $i=0,\ldots,N-1$, hold this Jacobian fixed over the horizon:
\begin{align}
        \tilde x_{i+1|k}
        =
        c_{D,k}
        +
        A_{D,k}\tilde x_{i|k}
        +
        B_{D,k}\tilde u_{i|k}.
        \label{eq:frozen_affine_model}
\end{align}

Define the estimated trim inputs
\begin{align}
        \hat u_{\rm trim,k}
        &\coloneqq
        \begin{bmatrix}
        \hat f_{\rm trim,k} & 0 & 0 & 0
        \end{bmatrix}^\top ,
        \\
        \hat{\tilde u}_{\rm trim,k}
        &\coloneqq
        \hat u_{\rm trim,k}-u_{\rme,0}.
        \label{eq:estimated_trim_inputs}
\end{align}
Parameterize the predicted input by its deviation from trim:
\begin{align}
        \tilde u_{i|k}
        =
        \hat{\tilde u}_{\rm trim,k}
        +
        \delta u_{i|k}.
        \label{eq:trim_parameterization}
\end{align}
Substitution into \eqref{eq:frozen_affine_model} gives
\begin{align}
        \tilde x_{i+1|k}
        =
        d_{D,k}
        +
        A_{D,k}\tilde x_{i|k}
        +
        B_{D,k}\delta u_{i|k},
        \label{eq:trimmed_predictor}
\end{align}
where
$d_{D,k} \coloneqq c_{D,k}
        + B_{D,k}\hat{\tilde u}_{\rm trim,k}$.

Following the PCAC timing convention, initialize
\begin{align}
        \tilde x_{0|k}
        &=
        \tilde x_k,
        \\
        \delta u_{0|k}
        &=
        u_k-\hat u_{\rm trim,k},
\end{align}
and define
\begin{align}
        \Delta\delta u_{i|k}
        \coloneqq
        \delta u_{i|k}-\delta u_{i-1|k},
        \qquad
        i=1,\ldots,N-1.
\end{align}

Lift \eqref{eq:trimmed_predictor} over the horizon:
\begin{align}
        X_{1|k}
        =
        \Gamma_{D,k}\tilde x_k
        +
        T_{D,k}U_{1|k}
        +
        C_{D,k},
        \label{eq:lifted_trimmed_predictor}
\end{align}
where
\begin{align}
        X_{1|k}
        &\coloneqq
        \operatorname{col}
        \left(
        \tilde x_{1|k},\ldots,\tilde x_{N|k}
        \right),
        \\
        U_{1|k}
        &\coloneqq
        \operatorname{col}
        \left(
        \delta u_{1|k},\ldots,\delta u_{N-1|k}
        \right),
        \\
        \Delta U_{1|k}
        &\coloneqq
        \operatorname{col}
        \left(
        \Delta\delta u_{1|k},\ldots,
        \Delta\delta u_{N-1|k}
        \right),
        \\
        E_{1|k}
        &\coloneqq
        \operatorname{col}
        \left(
        \varepsilon_{1|k},\ldots,\varepsilon_{N|k}
        \right),
        \\
        \hat U_{\rm trim,k}
        &\coloneqq
        \operatorname{col}
        \left(
        \hat u_{\rm trim,k},\ldots,
        \hat u_{\rm trim,k}
        \right).
        \label{eq:stacked_pcac_variables}
\end{align}
The lifted matrices $\Gamma_{D,k}$, $T_{D,k}$, and $C_{D,k}$ follow
\cite{nguyen2021predictive}.

For shifted reference $\tilde r_{i|k}$, solve
\begin{align}
\underset{U_{1|k},\,E_{1|k}}
{\operatorname{min}}
\quad
&
\frac{1}{2}
\sum_{i=1}^{N}
\left\|
\tilde x_{i|k}-\tilde r_{i|k}
\right\|_{Q_i}^{2}
\nonumber\\
&
+
\frac{1}{2}
\sum_{i=1}^{N-1}
\left\|
\Delta\delta u_{i|k}
\right\|_{R}^{2}
+
\frac{1}{2}
\sum_{i=1}^{N}
\left\|
\varepsilon_{i|k}
\right\|_{S}^{2}
\label{eq:pcac_qp}
\\
\operatorname{s.t.}
\quad
&
X_{1|k}
=
\Gamma_{D,k}\tilde x_k
+
T_{D,k}U_{1|k}
+
C_{D,k},
\nonumber\\
&
\mathcal H_xX_{1|k}
\leq
h_x+E_{1|k},
\nonumber\\
&
\mathcal U_{\min}
\leq
\hat U_{\rm trim,k}+U_{1|k}
\leq
\mathcal U_{\max},
\nonumber\\
&
\Delta\mathcal U_{\min}
\leq
\Delta U_{1|k}
\leq
\Delta\mathcal U_{\max},
\nonumber\\
&
E_{1|k}\geq0,
\nonumber
\end{align}
where $Q_i\succeq0$, $R\succ0$, and $S\succ0$; the constraint notation
follows \cite{nguyen2021predictive}.

Apply the first optimized deviation at step $k+1$:
\begin{align}
        u_{k+1}
        =
        \hat u_{\rm trim,k}
        +
        \delta u_{1|k}^{\star}.
        \label{eq:implemented_control}
\end{align}

\section{Numerical Evaluation}\label{sec:8}

\subsection{Simulation Setup}

The exact nonlinear plant \eqref{eq:sys_exact1}--\eqref{eq:sys_exact2} is
integrated by fixed-step RK4 with 10 substeps per sampling interval and
zero-order-held input. Twelve scalar RLS estimators, initialized from the
exact zero-order-hold discretization of the nominal hover-linear model,
identify the sampled-data map and share the VRF factor
\cite{bruce2020convergence}. At each sample, the identified map is
linearized and the PCAC problem is solved by the \texttt{quadprog}
active-set algorithm with at most 180 iterations and constraint tolerance
$10^{-7}$.

The public MATLAB implementation reproduces the reported cases through
\texttt{main\_pcac\_quadrotor.m}; all simulation and controller parameters
are centralized in \texttt{pcac\_parameters.m}, the reference trajectories
are defined in \texttt{generate\_reference.m}, and the reported metrics are
computed in \texttt{performance\_metrics.m}.

Case~1 imposes an abrupt mass change to test the mass and trim estimates.
Case~2 tracks an aggressive three-dimensional helix at constant mass to
test the higher-order predictors away from hover. Both cases use
$x(0)=0_{12}$, $u(0)=[m_0g,0,0,0]^\top$, noise-free full-state measurements,
and zero applied force and torque disturbances. All Taylor degrees use the
same plant, tuning, constraints, and horizon. Table~\ref{tab:simulation}
lists the parameters.

\begin{table}[t]
\centering
\caption{Simulation parameters.}
\label{tab:simulation}
\scriptsize
\setlength{\tabcolsep}{3pt}
\renewcommand{\arraystretch}{0.98}
\begin{tabular}{@{}ll@{}}
\toprule
Parameter & Value \\
\midrule
$T_s$ & $0.10~\mathrm{s}$ \\
$N$ & $10$ \\
$D$ & $\{1,2,3\}$ \\
\midrule
$m_0$ & $4.34~\mathrm{kg}$ \\
$g$ & $9.81~\mathrm{m/s^2}$ \\
$J$ & $\operatorname{diag}(0.0820,0.0845,0.1377)~\mathrm{kg\,m^2}$ \\
$m_0g$ & $42.5754~\mathrm{N}$ \\
\midrule
$u_{\min}$ & $[0,-2,-2,-1.2]^\top$ \\
$u_{\max}$ & $[85,2,2,1.2]^\top$ \\
$\Delta u_{\max}$ & $[6,0.35,0.35,0.25]^\top$ \\
$\Delta u_{\min}$ & $-\Delta u_{\max}$ \\
Attitude bounds & $|\psi|\leq\pi$, $|\phi|,|\theta|\leq50^\circ$ \\
\midrule
$Q$ & $\operatorname{diag}(110,110,150,2,3,3,$ \\
& $\qquad 18,18,28,0.7,0.7,0.7)$ \\
$Q_N$ & $8Q$ \\
$R$ & $\operatorname{diag}(0.015,0.18,0.18,0.12)$ \\
$S$ & $10^5I_6$ \\
\midrule
$P_{q,D,0}$ & $0.1I$ \\
$(\tau_n,\tau_d)$ & $(5,25)$ \\
$\eta$ & $0.50$ \\
Mass projection & $[2.5,7.0]~\mathrm{kg}$ \\
\bottomrule
\end{tabular}
\end{table}

The first component of each input-bound vector is thrust in newtons; the
remaining components are torques in newton-meters. Both cases use the
common 3-s reference ramp
\begin{equation}
s(t)=
\begin{cases}
0, & t\leq0,\\
\frac12\left[1-\cos\left(\frac{\pi t}{3}\right)\right], & 0<t<3,\\
1, & t\geq3.
\end{cases}
\end{equation}
In both cases, the velocity reference is the analytical derivative of the
position reference, and the attitude and angular-rate references are zero.
For $K$ samples, Tables~\ref{tab:mass_results}--\ref{tab:helix_results} use
\begin{align}
\mathrm{RMSE}_p
&=\sqrt{\frac1K\sum_{k=1}^{K}\|p_k-r_{p,k}\|_2^2},\\
\mathrm{RMSE}_{\mathrm{pred}}
&=\sqrt{\frac1{K-1}\sum_{k=2}^{K}\|e_{p,k}\|_2^2},\\
\mathrm{Input\ variation}
&=\sum_{k=1}^{K-1}\|u_{k+1}-u_k\|_1,\\
\mathrm{Maximum\ tracking\ error}
&=\max_k\|p_k-r_{p,k}\|_2,
\end{align}
where $e_{p,k}$ denotes the position components of the one-step prediction
error.

\subsection{Case 1: Abrupt Mass Change}

\begin{figure}[t]
    \centering
    \includegraphics[width=\columnwidth]{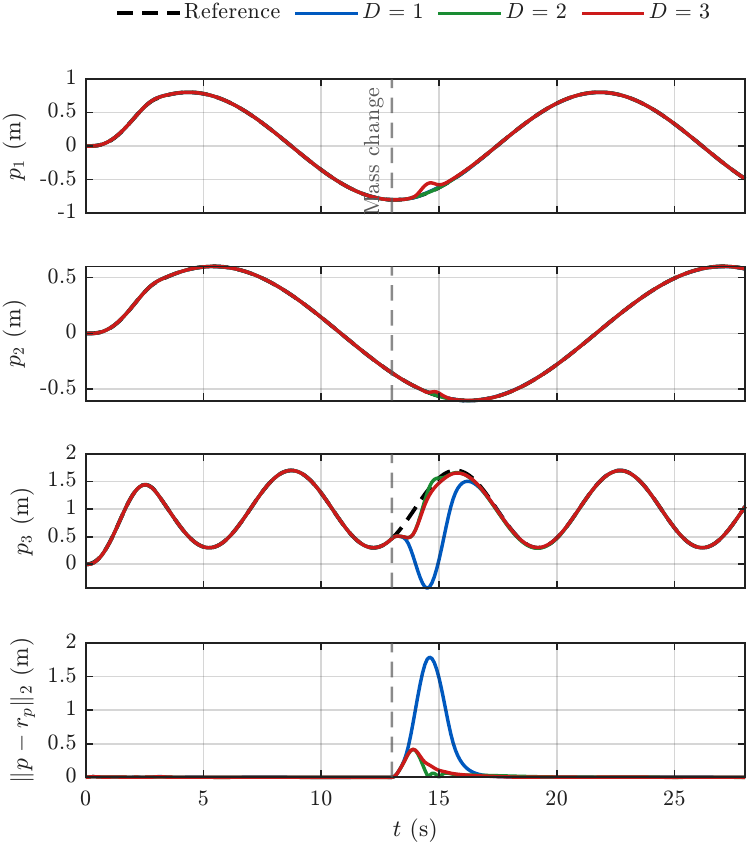}
    \caption{\textbf{Case 1}. Position tracking after the mass change at
    $t=13$ s; the lower panel shows $\lVert p-r\rVert_2$.}
    \label{fig:case1_tracking}
\end{figure}

\begin{figure}[t]
    \centering
    \includegraphics[width=\columnwidth]{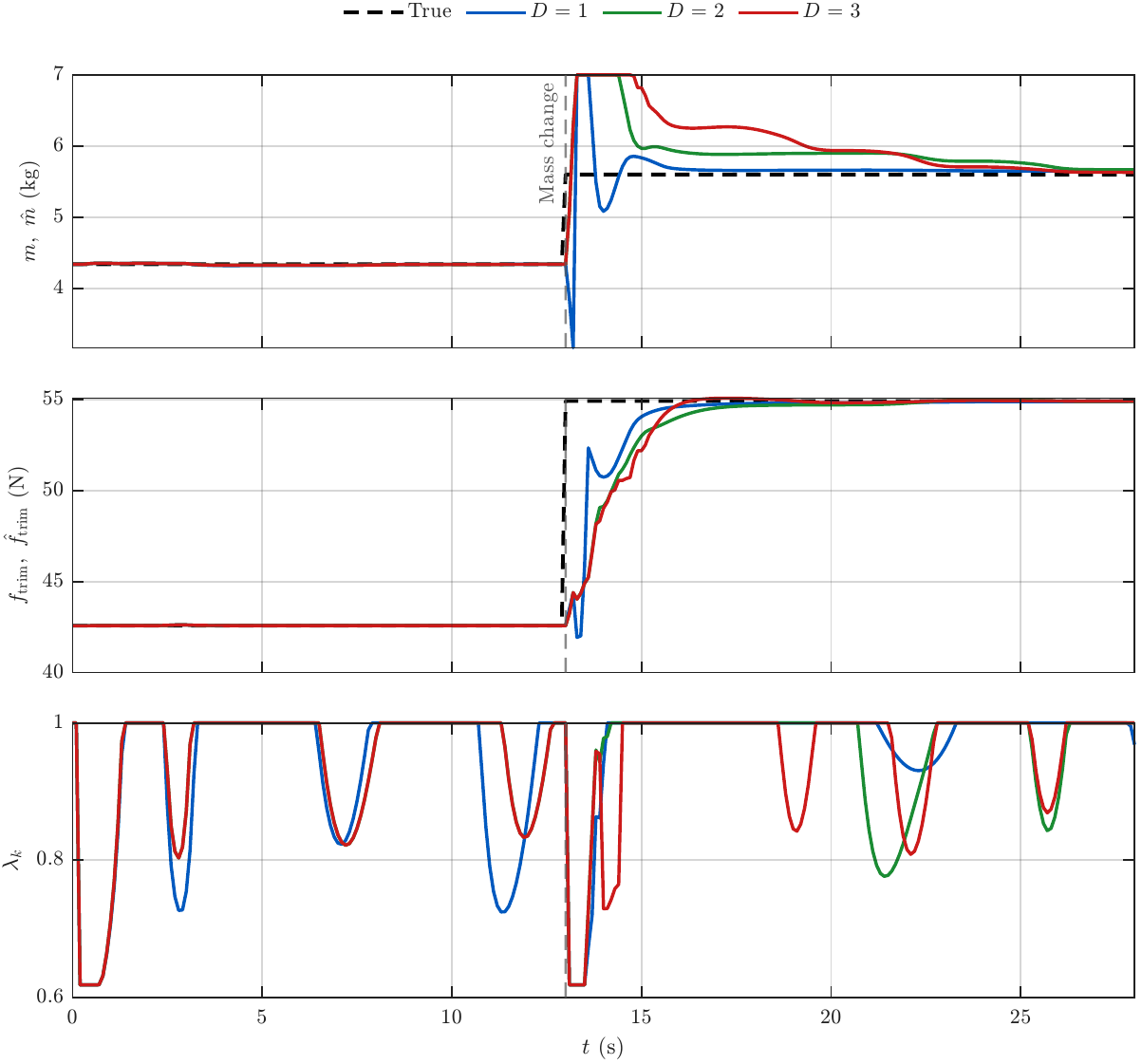}
    \caption{\textbf{Case 1}. Mass and trim-thrust estimates and VRF factor
    after the $4.34$--$5.60$ kg mass change. Dashed curves denote true values.}
    \label{fig:case1_estimates}
\end{figure}

The case ends at $t=28$ s. Its position reference is
\begin{align}
r_{p,1}&=0.80s(t)\sin(0.36t),\\
r_{p,2}&=0.60s(t)\sin(0.29t),\\
r_{p,3}&=s(t)\left[1+0.70\sin(0.90t)\right].
\end{align}
The mass changes from $m_0=4.34$ kg to $m=5.60$ kg at $t=13$ s.
Figs.~\ref{fig:case1_tracking} and \ref{fig:case1_estimates} show the
tracking response, online estimates, and VRF factor;
Table~\ref{tab:mass_results} reports the errors.

Both higher-order predictors reduce the tracking error. Degree two gives
the lowest position RMSE, whereas degree three gives the most accurate
terminal mass and trim estimates.

\begin{table}[t]
\centering
\caption{Case 1: Abrupt mass-change performance ($4.34\rightarrow5.60$ kg).}
\label{tab:mass_results}
\scriptsize
\setlength{\tabcolsep}{3pt}
\renewcommand{\arraystretch}{1.02}
\begin{tabular}{lccc}
\toprule
$D$ &
Pos. RMSE (m) &
Final $|m-\hat m|$ (kg) &
Final $|f_{\rm trim}-\hat f_{\rm trim}|$ (N)
\\
\midrule
$1$ & 0.348 & 0.052 & 0.028 \\
$2$ & \textbf{0.062} & 0.066 & 0.017 \\
$3$ & 0.069 & \textbf{0.031} & \textbf{0.007} \\
\bottomrule
\end{tabular}
\end{table}

\subsection{Case 2: Aggressive Helix Tracking}

\begin{figure}[t]
    \centering
    \includegraphics[width=\columnwidth]{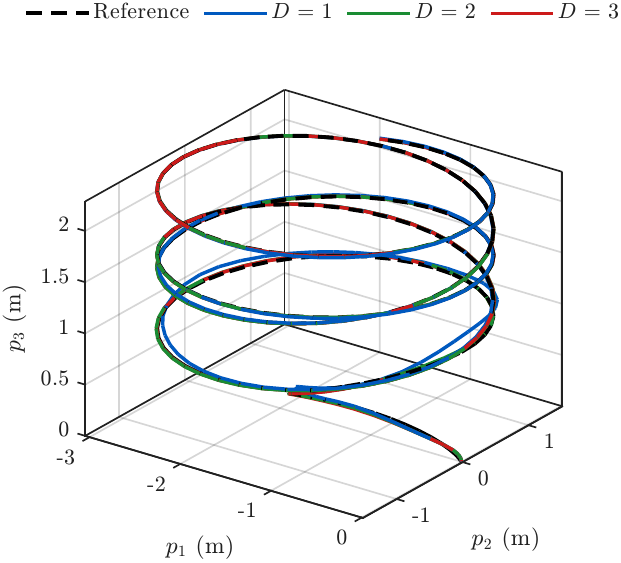}
    \caption{\textbf{Case 2}. Helical reference and closed-loop trajectories.}
    \label{fig:case2_trajectory}
\end{figure}

\begin{figure}[t]
    \centering
    \includegraphics[width=\columnwidth]{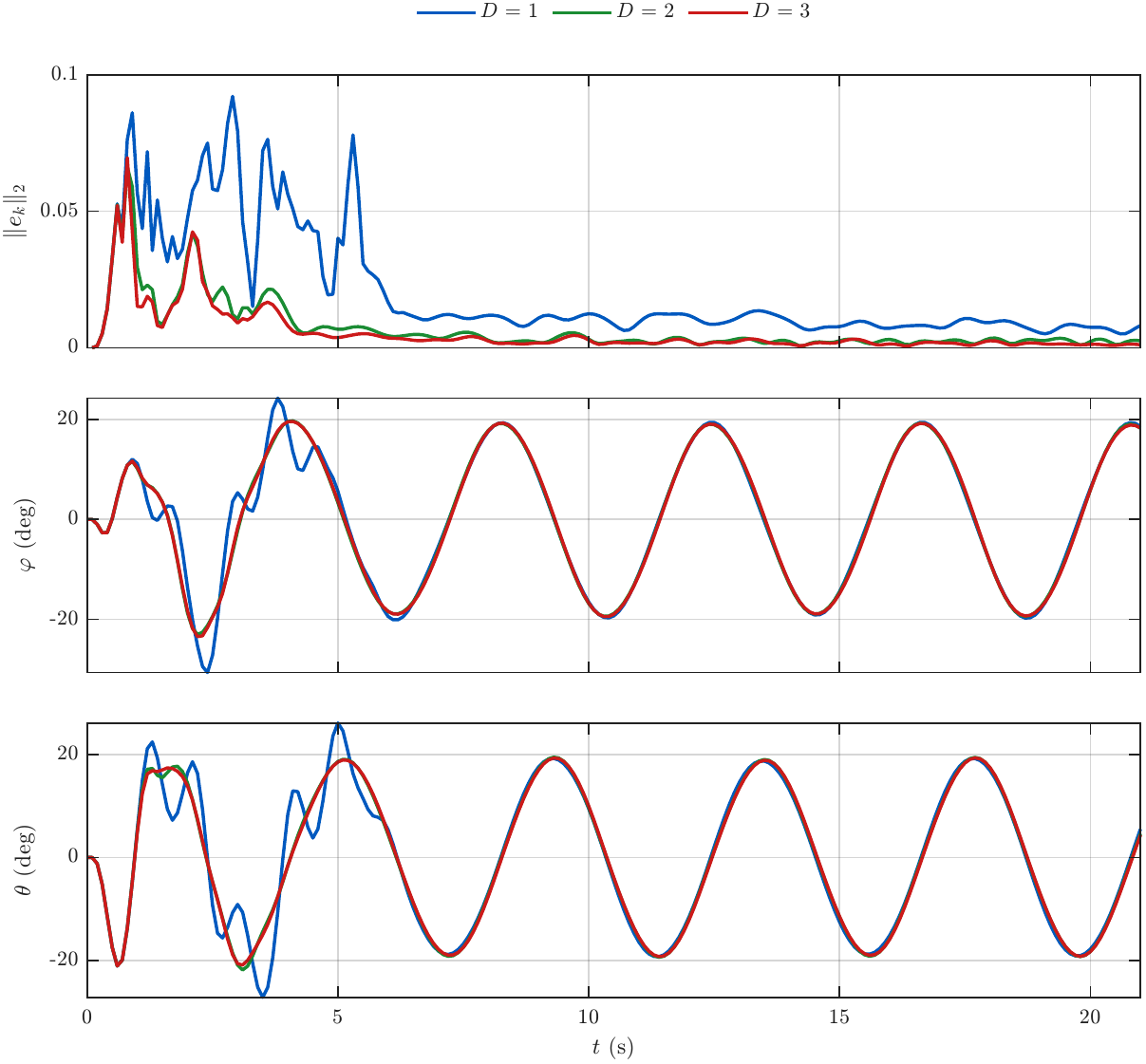}
    \caption{\textbf{Case 2}. One-step prediction-error norm and attitude
    responses during helix tracking.}
    \label{fig:case2_prediction_attitude}
\end{figure}

The quadrotor tracks a three-dimensional helix at constant mass $m=m_0=4.34$ kg until
$t=21$ s, with
\begin{align}
\chi(t)&=1.50(t-3),\\
r_{p,1}&=1.50s(t)[\cos\chi(t)-1],\\
r_{p,2}&=1.50s(t)\sin\chi(t),\\
r_{p,3}&=0.80s(t)+0.075\int_0^t s(\sigma)\,d\sigma\nonumber\\
&\quad+0.22s(t)\sin(0.78t).
\end{align}
Figure~\ref{fig:case2_trajectory} shows the trajectories, while
Fig.~\ref{fig:case2_prediction_attitude} shows the one-step prediction error
and attitude. Table~\ref{tab:helix_results} reports the performance metrics.

Degrees two and three provide similar tracking accuracy, but degree three
gives the lowest one-step prediction error and control variation.

\begin{table}[t]
\centering
\caption{Case 2: Aggressive helix tracking.}
\label{tab:helix_results}
\scriptsize
\setlength{\tabcolsep}{1.5pt}
\renewcommand{\arraystretch}{1.02}
\begin{tabular}{lcccc}
\toprule
$D$ &
Pos. RMSE (m) &
Pred. RMSE (m) &
Input var. &
Max. track. err. (m)
\\
\midrule
$1$ & 0.0519 & 0.001281 & 154.73 & 0.1695 \\
$2$ & 0.0288 & 0.000559 & 69.27 & 0.1664 \\
$3$ & \textbf{0.0281} & \textbf{0.000483} &
\textbf{59.41} & \textbf{0.1664} \\
\bottomrule
\end{tabular}
\end{table}

\FloatBarrier

\section{Conclusion}\label{sec:conclusion}

Taylor-informed PCAC combines sparse higher-order predictors with online
identification and gravity-trim adaptation. The second- and third-order
models improved tracking and one-step prediction relative to the first-order
model, while the identified vertical dynamics recovered the mass and trim
input after an abrupt payload change. These results show that local
higher-order structure can improve PCAC away from hover without changing
its optimization framework.

Future work will address motion-capture experiments and globally
nonsingular attitude representations.

\section*{Acknowledgment}

This research was partially supported by the MEXT SPReAD AI for Science program 2026 (Grant No. 26276994).

\bibliographystyle{IEEEtran}
\bibliography{bibliography}

@article{nguyen2021predictive,
  author  = {Nguyen, Tam W. and Islam, Syed Aseem Ul and Bernstein, Dennis S. and Kolmanovsky, Ilya V.},
  title   = {Predictive Cost Adaptive Control: A Numerical Investigation of Persistency, Consistency, and Exigency},
  journal = {IEEE Control Systems Magazine},
  year    = {2021},
  volume  = {41},
  number  = {6},
  pages   = {64--96},
  doi     = {10.1109/MCS.2021.3107647}
}

@misc{nguyen2026adaptivebehavioralpredictivecontrol,
  author        = {Nguyen, Tam W.},
  title         = {Adaptive Behavioral Predictive Control: State-Free Regulation Without Hankel Weights},
  year          = {2026},
  eprint        = {2602.12016},
  archivePrefix = {arXiv},
  primaryClass  = {eess.SY},
  url           = {https://arxiv.org/abs/2602.12016}
}

@misc{alhazmi2026nonlinearpredictivecostadaptive,
  author        = {Alhazmi, Rami Abdulelah and Babu, Achinth Suresh and Islam, Syed Aseem Ul and Bernstein, Dennis S.},
  title         = {Nonlinear Predictive Cost Adaptive Control of Pseudo-Linear Input-Output Models Using Polynomial, {F}ourier, and Cubic Spline Observables},
  year          = {2026},
  eprint        = {2602.05263},
  archivePrefix = {arXiv},
  primaryClass  = {eess.SY},
  url           = {https://arxiv.org/abs/2602.05263}
}

@article{bruce2020convergence,
  author  = {Bruce, Adam L. and Goel, Ankit and Bernstein, Dennis S.},
  title   = {Convergence and Consistency of Recursive Least Squares with Variable-Rate Forgetting},
  journal = {Automatica},
  year    = {2020},
  volume  = {119},
  pages   = {109052},
  doi     = {10.1016/j.automatica.2020.109052}
}

@inproceedings{nguyen2026fast,
  author       = {Nguyen, Tam W.},
  title        = {Fast {RLS} Identification Leveraging the Linearized System Sparsity: Predictive Cost Adaptive Control for Quadrotors},
  booktitle    = {2026 American Control Conference (ACC)},
  year         = {2026},
  organization = {IEEE}
}

@inproceedings{lee2010geometric,
  author       = {Lee, Taeyoung and Leok, Melvin and McClamroch, N. Harris},
  title        = {Geometric Tracking Control of a Quadrotor {UAV} on ${SE}(3)$},
  booktitle    = {49th IEEE Conference on Decision and Control (CDC)},
  year         = {2010},
  pages        = {5420--5425},
  organization = {IEEE},
  doi          = {10.1109/CDC.2010.5717652}
}

@article{dydek2013adaptive,
  author  = {Dydek, Zachary T. and Annaswamy, Anuradha M. and Lavretsky, Eugene},
  title   = {Adaptive Control of Quadrotor {UAV}s: A Design Trade Study with Flight Evaluations},
  journal = {IEEE Transactions on Control Systems Technology},
  year    = {2013},
  volume  = {21},
  number  = {4},
  pages   = {1400--1406},
  doi     = {10.1109/TCST.2012.2200104}
}

@inproceedings{hermand2018constrained,
  author       = {Hermand, Elie and Nguyen, Tam W. and Hosseinzadeh, Mehdi and Garone, Emanuele},
  title        = {Constrained Control of {UAV}s in Geofencing Applications},
  booktitle    = {2018 26th Mediterranean Conference on Control and Automation (MED)},
  year         = {2018},
  pages        = {217--222},
  organization = {IEEE},
  doi          = {10.1109/MED.2018.8443035}
}

@inproceedings{richards2025experimental,
  author       = {Richards, Riley J. and Marshall, Julius A. and Bernstein, Dennis S.},
  title        = {Experimental Flight Testing a Quadcopter Autopilot Based on Predictive Cost Adaptive Control},
  booktitle    = {2025 American Control Conference (ACC)},
  year         = {2025},
  pages        = {2471--2476},
  organization = {IEEE},
  doi          = {10.23919/ACC63710.2025.11108060}
}

@article{rastgoftar2021safe,
  author  = {Rastgoftar, Hossein and Kolmanovsky, Ilya V.},
  title   = {Safe Affine Transformation-Based Guidance of a Large-Scale Multiquadcopter System},
  journal = {IEEE Transactions on Control of Network Systems},
  year    = {2021},
  volume  = {8},
  number  = {2},
  pages   = {640--653},
  doi     = {10.1109/TCNS.2021.3084038}
}

@article{vanderschaaf2026active,
  author  = {Vander Schaaf, Jacob C. and Fidkowski, Krzysztof J. and Bernstein, Dennis S.},
  title   = {Active Flow Control Using Adaptive Model Predictive Control with Online, Closed-Loop System Identification},
  journal = {Journal of Guidance, Control, and Dynamics},
  year    = {2026},
  volume  = {49},
  number  = {7},
  pages   = {1883--1896}
}

@article{he2026projection,
  author  = {He, Tianyi and Burton, Samantha and Spencer, Clayton and Wei, Wenpeng},
  title   = {Projection-Based Online Parameter Estimation of a Tilt-Rotor {VTOL} Aircraft and Experimental Validation},
  journal = {ASME Letters in Dynamic Systems and Control},
  year    = {2026},
  volume  = {6},
  number  = {2},
  pages   = {021012},
  doi     = {10.1115/1.4070611}
}

@article{clarke1987generalized,
  author  = {Clarke, David W. and Mohtadi, Coorous and Tuffs, P. Simon},
  title   = {Generalized Predictive Control---Part {I}: The Basic Algorithm},
  journal = {Automatica},
  year    = {1987},
  volume  = {23},
  number  = {2},
  pages   = {137--148},
  doi     = {10.1016/0005-1098(87)90087-2}
}

@article{clarke1987generalized2,
  author  = {Clarke, David W. and Mohtadi, Coorous and Tuffs, P. Simon},
  title   = {Generalized Predictive Control---Part {II}: Extensions and Interpretations},
  journal = {Automatica},
  year    = {1987},
  volume  = {23},
  number  = {2},
  pages   = {149--160},
  doi     = {10.1016/0005-1098(87)90088-4}
}

@article{tao2026robust,
  author  = {Tao, Ran and Zhao, Pan and Kolmanovsky, Ilya V. and Hovakimyan, Naira},
  title   = {Robust Adaptive {MPC} in the Presence of Nonlinear Time-Varying Uncertainties: An Uncertainty Compensation Approach},
  journal = {Automatica},
  year    = {2026},
  volume  = {192},
  pages   = {113142},
  doi     = {10.1016/j.automatica.2026.113142}
}

@article{richards2025benchmark,
  author  = {Richards, Riley J. and Islam, Syed Aseem Ul and Bernstein, Dennis S.},
  title   = {Predictive Cost Adaptive Control of the {NASA} Benchmark Flutter Model},
  journal = {Journal of Guidance, Control, and Dynamics},
  year    = {2025},
  volume  = {48},
  number  = {12},
  pages   = {2663--2679},
  doi     = {10.2514/1.G008859}
}

@inproceedings{farahmandi2024missile,
  author       = {Farahmandi, Alireza and Reitz, Brian},
  title        = {Predictive Cost Adaptive Control of a Planar Missile with Unmodeled Aerodynamics},
  booktitle    = {AIAA SCITECH 2024 Forum},
  year         = {2024},
  organization = {AIAA},
  doi          = {10.2514/6.2024-2218}
}

@inproceedings{lai2024adaptive,
  author       = {Lai, Brian and Bernstein, Dennis S.},
  title        = {Adaptive {Kalman} Filtering Developed from Recursive Least Squares Forgetting Algorithms},
  booktitle    = {2024 American Control Conference (ACC)},
  year         = {2024},
  pages        = {4378--4383},
  organization = {IEEE},
  doi          = {10.23919/ACC60939.2024.10644929}
}

@article{burtsev2026adaptive,
  author  = {Burtsev, Anton and Jariwala, Akshit and Bakolas, Efstathios and Goldstein, David},
  title   = {Adaptive Feedback Flow Control for Wings},
  journal = {AIAA Journal},
  year    = {2026},
  pages   = {1--14},
  doi     = {10.2514/1.J066619}
}

@article{huang2026fls,
  author  = {Huang, Zhengguo and Chen, Mou and Shen, Hao},
  title   = {{FLS}-Based Adaptive Flight Control of Fixed-Wing {UAV} with Augmented Switching Model},
  journal = {IEEE Transactions on Systems, Man, and Cybernetics: Systems},
  year    = {2026},
  volume  = {56},
  number  = {3},
  pages   = {2036--2048},
  doi     = {10.1109/TSMC.2025.3649695}
}

@article{eren2017model,
  author  = {Eren, Utku and Prach, Anna and Ko\c{c}er, Ba\c{s}aran Bahad\i{}r and Rakovi\'{c}, Sa\v{s}a V. and Kayacan, Erdal and A\c{c}\i{}kme\c{s}e, Beh\c{c}et},
  title   = {Model Predictive Control in Aerospace Systems: Current State and Opportunities},
  journal = {Journal of Guidance, Control, and Dynamics},
  year    = {2017},
  volume  = {40},
  number  = {7},
  pages   = {1541--1566},
  doi     = {10.2514/1.G002507}
}

\end{document}